\documentclass[%
 reprint,
 amsmath,amssymb,
 aps, prl
]{revtex4-2}
\usepackage{graphicx}
\usepackage{textcomp}
\usepackage{amssymb}
\usepackage[colorlinks,linkcolor=blue,anchorcolor=blue,citecolor=blue,urlcolor=black]{hyperref}
\usepackage[utf8]{inputenc}
\usepackage{amsmath}
\usepackage{tipa}
\usepackage{comment}
\usepackage{mathtools}
\usepackage{braket}
\usepackage{listings}
\usepackage{xcolor}
\usepackage{soul}
\usepackage{etoolbox}
\usepackage{lineno}
\usepackage{ulem}
\usepackage{svg}
\usepackage{CJKutf8}

\renewcommand{\vec}[1]{\mathbf{{#1}}}
\definecolor{Darkgreen}{rgb}{0,0.4,0}
\definecolor{listinggray}{gray}{0.9}
\definecolor{lbcolor}{rgb}{0.9,0.9,0.9}
\begin{document}

\title{Fringe visibility and which-way information in Young’s double slit experiments with light scattered from single atoms
} 

\author{Hanzhen Lin}
\affiliation{Department of Physics, Massachusetts Institute of Technology, Cambridge, MA 02139, USA}
\affiliation{Research Laboratory of Electronics, Massachusetts Institute of Technology, Cambridge, MA 02139, USA}
\affiliation{MIT-Harvard Center for Ultracold Atoms, Cambridge, MA, USA}

\author{{Yu-Kun Lu}}
\affiliation{Department of Physics, Massachusetts Institute of Technology, Cambridge, MA 02139, USA}
\affiliation{Research Laboratory of Electronics, Massachusetts Institute of Technology, Cambridge, MA 02139, USA}
\affiliation{MIT-Harvard Center for Ultracold Atoms, Cambridge, MA, USA}

\author{{Vitaly Fedoseev}}
\affiliation{Department of Physics, Massachusetts Institute of Technology, Cambridge, MA 02139, USA}
\affiliation{Research Laboratory of Electronics, Massachusetts Institute of Technology, Cambridge, MA 02139, USA}
\affiliation{MIT-Harvard Center for Ultracold Atoms, Cambridge, MA, USA}

\author{Yoo Kyung Lee}
\affiliation{Department of Physics, Massachusetts Institute of Technology, Cambridge, MA 02139, USA}
\affiliation{Research Laboratory of Electronics, Massachusetts Institute of Technology, Cambridge, MA 02139, USA}
\affiliation{MIT-Harvard Center for Ultracold Atoms, Cambridge, MA, USA}

\author{{Jiahao Lyu}}
\affiliation{Department of Physics, Massachusetts Institute of Technology, Cambridge, MA 02139, USA}
\affiliation{Research Laboratory of Electronics, Massachusetts Institute of Technology, Cambridge, MA 02139, USA}
\affiliation{MIT-Harvard Center for Ultracold Atoms, Cambridge, MA, USA}

\author{Wolfgang Ketterle}
\affiliation{Department of Physics, Massachusetts Institute of Technology, Cambridge, MA 02139, USA}
\affiliation{Research Laboratory of Electronics, Massachusetts Institute of Technology, Cambridge, MA 02139, USA}
\affiliation{MIT-Harvard Center for Ultracold Atoms, Cambridge, MA, USA}
%
    
\maketitle
\textbf{
Young’s double slit experiment has often been used to illustrate the concept of complementarity in quantum mechanics. If information can in principle be obtained about the path of the photon, then the visibility of the interference fringes is reduced or even destroyed. This Gedanken experiment discussed by Bohr and Einstein can be realized when the slit is replaced by individual atoms sensitive to the transferred recoil momentum of a photon which ``passes through the slit''\cite{bohr_chapter,bohr_collected_work,jammer_2015_philosophy_QM,fedoseev2024coherent,hefei2024}.
Early pioneering experiments were done with trapped ions and atom pairs created via photo-dissociation \cite{wineland1998_two_atom_fringe, eichmann_two_ion_fringe_1993,alain_aspect_calcium2_recoil_interference_1982}.  
Recently, it became possible to perform interference experiments with single neutral atoms cooled to the absolute ground state of a harmonic oscillator potential. The slits are now single atoms representing a two-level system, and the excitation in the harmonic oscillator potential is the which-way marker.  In this note, we analyze and generalize two recent experiments performed with single atoms \cite{fedoseev2024coherent,hefei2024} and emphasize the different ways they record which-way information. }

%
In the early development of quantum mechanics, several important thought experiments considered what would happen if photons were sent through different configurations of slits in order to illustrate the concepts of particle-wave duality and entanglement \cite{bohr_chapter,bohr_collected_work,zurek1979_complementarity,jammer_2015_philosophy_QM}.
In particular, Niels Bohr famously showed how the complementarity principle protects the consistency of quantum mechanics \cite{bohr_chapter,bohr_collected_work}. 
Certain measurements are mutually exclusive and demonstrate either the photon's wave nature (by recording the interference pattern) or its particle nature (by recording the path taken by the photon).

\begin{figure}
    \centering
    \includegraphics[width= \linewidth]{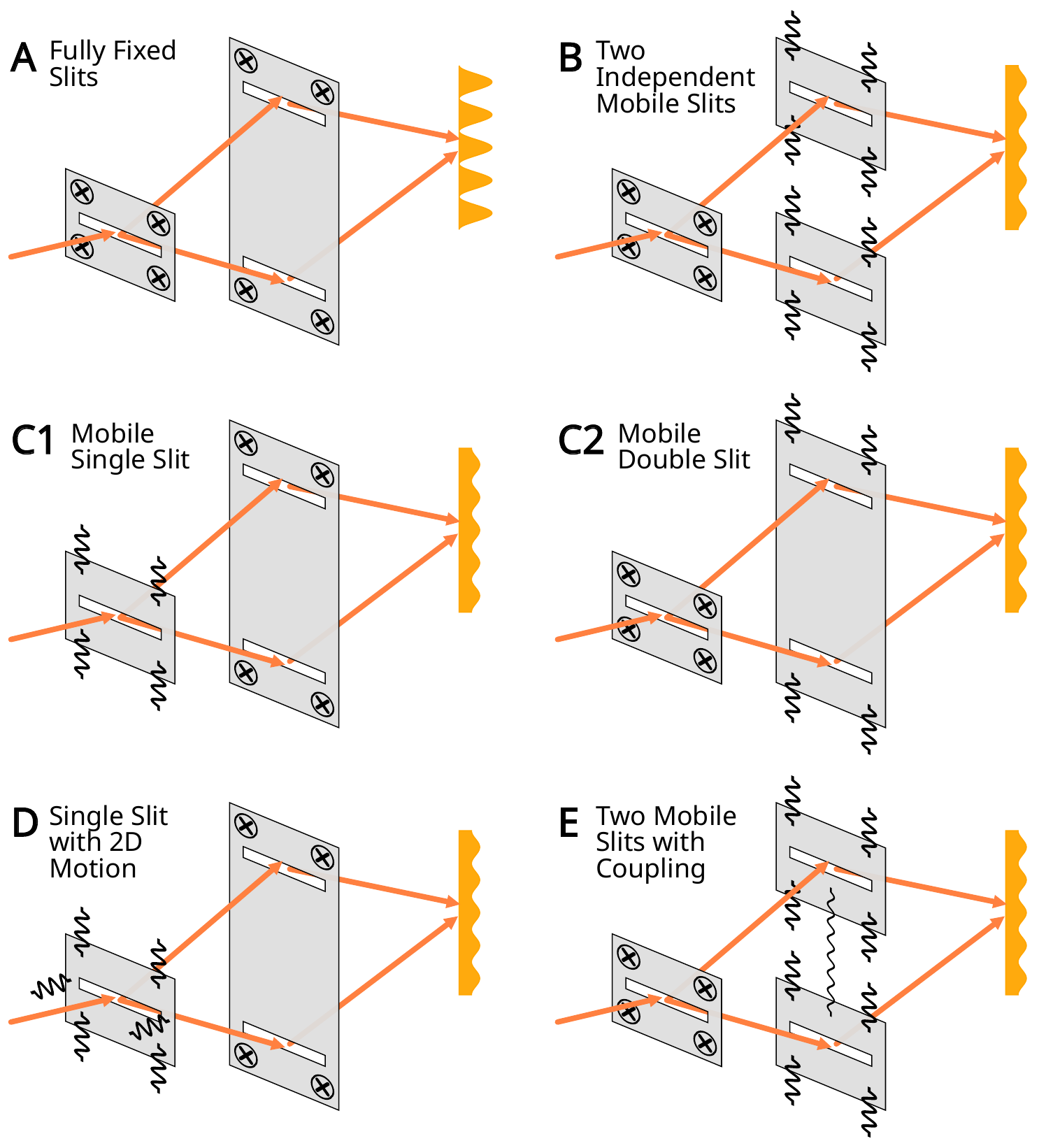}
    \caption{
    Different configurations for double-slit experiments where partial which-way information is obtained. 
    A rigid double-slit geometry  (\textbf{A}) creates full interference contrast (yellow). 
    The interference pattern has reduced contrast and a background when the slits are allowed to move, and we consider three possible configurations: in \textbf{B}, the two slits of the double slit move independently; in \textbf{C1} the single slit moves; in \textbf{C2} the double slit moves as one unit.
    These three cases can be generalized to allow motion longitudinal to the incident photon direction as well, as shown by \textbf{D}, which is  configuration \textbf{C1} with 2D motion. 
    In addition, configuration \textbf{E} is an intermediate configuration between \textbf{B} and \textbf{C2}, where two mobile slits are coupled by a weak spring.
   }
    \label{fig:interferometer}
\end{figure}

The classic double slit experiment has three slits: the first filters out a single transverse mode that coherently illuminates the double slit (Fig. \ref{fig:interferometer}A). 
To obtain information about the path taken by photons, one could imagine allowing the slits to move \cite{bohr_chapter,bohr_collected_work} (Fig. \ref{fig:interferometer} B and C) and observing the recoil experienced by the slits.
Einstein introduced configuration C1 in Fig. \ref{fig:interferometer} to question the consistency of quantum mechanics at the Fifth Solvay Congress in 1927 \cite{jammer_2015_philosophy_QM,zurek1979_complementarity}. 
This configuration has been analyzed in detail \cite{zurek1979_complementarity, utter2007_trapped_ion_recoiling_slit}. 
Since then, the paradigm of the Einstein-Bohr recoiling double slit has been discussed in numerous papers, sometimes adopting other variations \cite{haroche2001complementarity,liu2015_double_slit_molecular_oxygen,fedoseev2024coherent,hefei2024}. 
The arrangements C1 and C2 in Fig. \ref{fig:interferometer} (henceforth jointly referred to as case C) are equivalent and record the photon's path by the direction of motion of the slit(s) whereas in arrangement B, the path taken by the photon is recorded by checking which slit has started to move.
Both configurations B and C (as well as their generalizations D and E, to be discussed later) illustrate Bohr's argument that an accurate measurement of the recoil momentum of the slit that is sufficiently accurate to determine the path of the particle necessarily introduces an uncertainty in the position of the slit which will wash out the interference pattern.
In the limit where the masses of the slits are large, all configurations reduce to fully fixed slits (Fig.~\ref{fig:interferometer}A), where no which-way information is revealed; in the limit of small masses, the mobile slits measure which-way information for each photon fully, and no interference pattern is recorded.

In this short note, we compare the different geometries shown in Fig.~\ref{fig:interferometer} in the intermediate regime where partial which-way information is obtained while observing partial interference contrast. 
In particular, we show that a reduction in the fringe visibility is not necessarily due to the recording of which-way information. 
For example, the visibility reduction can occur when the which-way information is phase-sensitive and not robustly stored in an eigenstate of the slit.

In the intermediate regime, when which-way information is only partially obtained, the two recent experimental arrangements have a distinctive difference \cite{fedoseev2024coherent,hefei2024}. 
One of them observed the interference of light between many single atoms prepared in a Mott insulator state in an optical lattice \cite{fedoseev2024coherent}. 
This experiment is conceptually equivalent to a double-slit experiment where light is scattered by two atoms and partially interferes, realizing scheme B.
The other experiment, which used a single atom scattering light into two directions and observed partial interference, realizes the scheme C1 \cite{hefei2024}. 
The standard interpretation of such experiments is that a reduced contrast implies which-way information has been obtained. 
However, as we explain here, this is not always the case.
Although the two experiments are superficially similar --- two atoms scatter light into one direction, versus one atom scattering light into two directions --- the way they obtain and store which-way information is fundamentally different.

We highlight these differences by considering the regime of short and long light pulses for each experiment as well as the possibility of a so-called ``quantum eraser'' which can be used to erase which-way information and thereby re-establish full interference between the two pathways.
Although every part of our discussion can be found in textbooks or in the literature, we see additional value and clarity by directly comparing various configurations and protocols (short and long pulses, quantum eraser) using the same analysis. 
Therefore, we have developed a model where the recording and erasure of which-way information is realized with mechanical oscillators. This is equivalent to previous work where microwave emission or photon scattering were used as which-way markers\cite{scully1991_quantum_optic_test_complementarity,feynman_lec_vol3}. 

Without loss of generality, we assume the atoms (slits) are constrained to move in only one dimension unless specified. Each slit with mass $m$ is in an harmonic trap with frequency $1/\omega_{\rm trap}$, characteristic oscillator length $x_0=\sqrt{\hbar/m\omega_{\rm trap}}$, lowering operator $a$, and displacement operator from the center of each trap $\vec{R}=(a^\dagger+a)x_0/\sqrt{2} $.
Rayleigh scattering of a photon can be described by operator $D=\exp(iQR)$, which effectively displaces the momentum of an atom by the photon recoil $\hbar Q = \hbar(k_{\mathrm{in}} - k_{\mathrm{out}})$, where $k_{\mathrm{in},\mathrm{out}}$ are the wave-vector of the incident and scattered photon, projected along the direction where the slit can move. For simplicity, assume the the two path impart equal and opposite recoil to the atoms, which move the atoms from their ground state $\ket{0}$ to a momentum-displaced coherent state $\ket{\pm\beta}$ with $\beta=\frac{iQx_0}{\sqrt{2}}$.
In case B with two independent atoms, each atom scatters the photon into their respective path, and we denote the total state by $|\text{atom 1}, \text{atom 2}; \text{path 1}, \text{path 2} \rangle$, the atom state by $\ket{\cdots}_{\text{a}}$, and the photon state by $\ket{\cdots}_{\gamma}$.
After a short light pulse with duration $\ll 1/\omega_{\rm trap}$), the resulting total state is
\begin{equation}
|\psi\rangle = |0,0;0,0\rangle + \epsilon (\gamma_1 |\beta,0;1,0\rangle + \gamma_2 |0, -\beta;0, 1\rangle)
\end{equation}

Here $\epsilon\ll 1$ is a small parameter describing the probability of scattering a photon and $\gamma_{\{1,2\}}=\exp(iQR_{\{1,2\}})$ describes the scattering phase of each path. 
One can choose to set them both to 1, but the extra label helps identify the contributions from the two atoms.
%
%
For small $\beta$, we expand  $\ket{\pm\beta} \approx \ket{0}\pm\beta\ket{1}$ and obtain
\begin{align}
|\psi\rangle = &|0,0;0,0\rangle + \epsilon \beta \gamma_1 \ket{1,0;1,0} - \epsilon \beta \gamma_2 \ket{0,1;0,1} \nonumber \\
& +\epsilon \ket{0,0}_{\text{a}}\otimes(\gamma_1 \ket{1,0}_{\gamma}+ \gamma_2\ket{0,1}_{\gamma})
\label{eqn:perturbative_two_atom}
\end{align}

By examining the last three terms for which a photon has been scattered, we obtain relative probabilities of $|\beta|^2/2$ each for identifying that the photon has passed through the upper/lower slit, and $1 - |\beta|^2$ for recording no which-way information in which case and the two paths interfere on the detector with a symmetric pattern (expressed by  $\gamma_1 \ket{1,0}_{\gamma}+ \gamma_2\ket{0,1}_{\gamma}$).
 Therefore, the interference pattern for the scattered light has a contrast of $1-|\beta|^2$.

The recorded information can be erased by a quantum eraser \cite{scully1991_quantum_optic_test_complementarity}, which performs a $\pi/2$ rotation on the two atomic states $|1,0\rangle \rightarrow \frac{1}{\sqrt{2}}(|1,0\rangle + |0,1\rangle), \quad |0,1\rangle \rightarrow \frac{1}{\sqrt{2}}(|1,0\rangle - |0,1\rangle)$. 
Such an operation is experimentally possible for neutral atom using a two-photon Raman transition where one of the ``beams'' is a single photon in a cavity coupled to both atoms, or by using Rydberg interactions\cite{rempe_2018_cavity_mediated_gate,endres_2025_atom_moition_coupled_by_rydberg}. The quantum eraser prepares the state:
\begin{align}
|\psi\rangle &= \ket{0,0;0,0} + \epsilon \ket{0,0}_{\text{a}}\otimes(\gamma_1 \ket{1,0}_{\gamma}+ \gamma_2\ket{0,1}_{\gamma}) \nonumber \\
&\quad + \frac{\epsilon \beta}{\sqrt{2}} \ket{1,0}_{\text{a}} \otimes(\gamma_1 \ket{1,0}_{\gamma}- \gamma_2\ket{0,1}_{\gamma}) \nonumber \\
&\quad + \frac{\epsilon \beta}{\sqrt{2}} \ket{0,1}_{\text{a}} \otimes(\gamma_1 \ket{1,0}_{\gamma}+ \gamma_2\ket{0,1}_{\gamma})
\end{align}

Now the interference pattern has to be measured in coincidence with the atomic state, and for $\ket{0,1}_{\text{a}}$ the interference pattern is symmetric, i.e., it has the same phase as for the rigid slit; for $\ket{1,0}_{\text{a}}$ the interference pattern is $\pi$-shifted (or antisymmetric), expressed by the $\gamma_1 \ket{1,0}_{\gamma}- \gamma_2\ket{0,1}_{\gamma}$ term.
This implies that full interference contrast has been restored by the quantum eraser, and all which-way information has been erased. 
Note that the eraser operation is unitary and reversible, and it is only the measurement process which provides information about the photon's particle or wave aspect.

Alternatively, the whole experiment can be done with light pulses much longer than the period of the harmonic oscillator potential. 
In this case, one must use Fermi's golden rule and the system becomes a mixture. The possible states after scattering are $\ket{0,0}_{\text{a}}\otimes(\gamma_1 \ket{1,0}_{\gamma}+ \gamma_2\ket{0,1}_{\gamma})$ with shows interference, and two non-interfering states $\ket{1,0;1,0}$, $\ket{0,1;0,1}$. Probabilities of the three states are given by the squares of the amplitudes in Eq. \ref{eqn:perturbative_two_atom}.
At this point, information about the phase factors for the states $\ket{1,0;1,0}$ and $\ket{0,1;0,1}$ has been lost, and it is not possible to regain full interference contrast via a quantum eraser~\cite{footnote}.

Let us apply the same analysis to the configuration C where the momentum transfer of the photon is recorded by a single harmonic oscillator.
%
Denoting now the quantum states by $|\text{atom}; \text{path 1}, \text{path 2}\rangle$, after a short light pulse (much shorter than the harmonic oscillator period) the initial state $|0;0,0\rangle$ evolves to  
\begin{equation}
|\psi\rangle = |0;0,0\rangle + \epsilon (\gamma_1 |\beta;0,1\rangle + \gamma_2 |-\beta;1,0\rangle)
\end{equation}

The direction of the light scattering is now entangled with opposite momentum transferred to the atoms, leading to coherent states $|\pm \beta\rangle$. 
Since the coherent states $ \ket{+\beta}$ and $\ket{-\beta}$ are not orthogonal, it is impossible to obtain definitive which-way information, only probabilistic information. 
For instance, if the measurement is performed by projecting the final atomic state onto a coherent state $|\delta\rangle$, the probability for a positive measurement is $\exp(-|\delta\mp\beta|^2 )$ for $\ket{\pm \beta}$ states. 
The ratio of these probabilities implies that the path is identified with a fractional error $\exp(-4|\beta\delta|)$. 
However, a small error requires a small probability $\approx \exp(-|\delta|^2)$ to detect the recoil transferred to the atom. 
This interplay of certainty of information and probability of obtaining it has been discussed in detail \cite{zurek1979_complementarity,utter2007_trapped_ion_recoiling_slit}.

The major difference between configuration B with two atoms is that for small momentum transfers $(\beta < 1)$ the atomic Hilbert space is only two-dimensional, while it is three-dimensional for the independent slits. 
Expanding for small $\beta$ gives
\begin{align}    
|\psi\rangle &= \quad |0;0,0\rangle + \epsilon (\gamma_1 |0;1,0\rangle + \gamma_2 |0;0,1\rangle) \\
&\quad+ \epsilon \beta (\gamma_1 |1;1,0\rangle - \gamma_2 |1;0,1\rangle ) \\
&= |0;0,0\rangle+ \epsilon \ket{0}_{\text{a}}\otimes (\gamma_1 \ket{1,0}_{\gamma} + \gamma_2\ket{0,1}_{\gamma}) \nonumber \\
&\quad + \epsilon \beta \ket{1}_{\text{a}}\otimes(\gamma_1 \ket{1,0}_{\gamma} - \gamma_2\ket{0,1}_{\gamma})
\end{align}
If the interference between the two paths $ |1,0\rangle_{\gamma}$ and $|0,1\rangle_{\gamma}$ is measured in coincidence with $|0\rangle_\text{a}$ or $|1\rangle_\text{a}$, one observes full interference contrast with a symmetric or anti-symmetric phase. Without detecting the atomic state, the interference contrast is reduced to  $1 - 2 |\beta|^2$. 

In case of a long light pulse, Fermi’s golden rule gives relative probabilities of $1 - |\beta|^2$ and $|\beta|^2$  for a symmetric interference pattern (identical to case A with rigid slits), or a $\pi$-shifted pattern, respectively. 
The shifted pattern is accompanied by the excitation of the atomic state $|1\rangle$ and a frequency shift $\omega_{\rm trap}$ relative to the incident light. In contrast to the short pulses, unless the scattering time is precisely measured (much better than the trap period), the continuously evolving phase of the atomic exitation cannot be resolved. 
When all photons are recorded, the interference contrast is reduced to  $1 - 2 |\beta|^2$. 
However, the frequency shift can be used to obtain full-contrast interference for both frequencies. 
For instance, instead of measuring the interference pattern separately for the two frequencies, one could include a dispersive element which causes a relative phase shift of $\pi$ for the frequency shifted light, in which case a single detector would observe full contrast. 
Thus, in configuration C, which-way information can be recorded only via the phase between $|0\rangle$ and $|1\rangle$ of the superposition state which requires the temporal resolution of a short pulse. 
For long light pulses in which such phases are not recorded, the apparent loss of fringe contrast is \textit{not} related to which-way information, but reflects that the unshifted and $\pi$ phase-shifted interference patterns are entangled with different final atomic states $|0\rangle$ and $|1\rangle$. 
The recent single-atom experiment confirmed the reduction of interference contrast by $1 - 2 |\beta|^2$ due to such atom-photon entanglement \cite{hefei2024}. 
However which-way information with sufficient certainty can only be obtained with a probability smaller than $|\beta|^2$ \cite{zurek1979_complementarity}.

In configuration C, full contrast can be obtained if the photons are measured using a dispersive optical element or by recording photons in coincidence with the atomic state.
This is fundamentally different from configuration B, where the photon recoil transferred to the two separate slits \textit{always} records which-way information. 
Full contrast can only be obtained with a quantum eraser, i.e., by modifying the recorded information before it is read out.

To see this distinction more clearly, we can extend the discussion to allow the slits to move also in the longitudinal ($z$) direction (configuration D).
If a similar modification is made for configuration B, the added dimension will make no difference --- any excitation of a single-slit provides which-way information. 
However, when making such modification to configuration C, the longitudinal momentum transfer is common mode for the two paths and does not provide which-way information, as we derive below.

Scattering photons into the two directions creates a displacement of the harmonic oscillator ground state in the  $z$-direction to $|\alpha\rangle$ and in the $y$-direction to $|\pm \beta\rangle$. 
Denoting states by $|\text{atom}_z, \text{atom}_x; \text{path 1}, \text{path 2} \rangle$
one obtains the final state as
\begin{equation}
|\psi\rangle = |0,0;0,0\rangle + \epsilon |\alpha\rangle \otimes(\gamma_1 |\beta;0,1\rangle + \gamma_2 |-\beta;1,0\rangle)
\end{equation}
The displacement in $z$ is common mode for the interference between two paths and does not affect the interference pattern or contrast. 
This is valid even if the harmonic confinement along $z$ is much weaker than along $y$ and therefore $\alpha \gg 1$. 
In this case, every scattering event will excite the $z$-motion, and therefore can be detected with almost 100\% probability. 
But this does not provide any information regarding the path of the photon. 
In contrast, for the case of independent slits, any detected momentum transfer will provide which-way information and reduce the interference contrast.

There is an intermediate case between independent double slits  and connected double slits (Fig. \ref{fig:interferometer}B and C2), by allowing for a weak coupling between the motion of the two slits as depicted in Fig. \ref{fig:interferometer} E. 
In practice, such coupling can come from Coulomb interactions in ion traps, or a weak molecular bond between the two atoms \cite{liu2015_double_slit_molecular_oxygen,wolf_small_ion_crystal_fringe_2016}. 
As a consequence of this weak coupling, the eigenstates for the excitation are symmetric and anti-symmetric respectively, whose frequency differ by a small beat frequency. 
For light pulse shorter than the beat period, they behave like independent slits. 
The coupling of the two slits implements a quantum eraser after exactly a quarter of the beat period, and full interference contrast can be restored by a coincidence measurement in the basis of upper or lower slit excitations. 

For long light pulses (much longer than the beat note period), information is always recorded in energy eigenstates.
However, in contrast to case B where the which-way markers are the eigenstates, the eigenstates in E (and C) are \textit{not} which-way markers: instead, they are correlated with a symmetric or $\pi$ phase-shifted interference pattern.
Full interference contrast can now be regained by recording the contrast in coincidence with reading out the atomic states (i.e. symmetric or anti-symmetric states of the slits), or by inserting a suitable dispersive element into the light path which provides a relative phase shift of $\pi$ between the light shifted by the symmetric and anti-symmetric excitation frequencies. In this case which-way information has never been recorded since for light pulse duration much longer than the beat period, it is fundamentally no longer possible to infer from which slit the excitation originated.

In this note, we have discussed configurations where photons are scattered by slits realized by individual ions or atoms held in place by an harmonic oscillator potential, which has been realized in several recent and past experiments \cite{fedoseev2024coherent,wineland1998_two_atom_fringe,hefei2024,eichmann_two_ion_fringe_1993,wolf_small_ion_crystal_fringe_2016}. Related experiments have also been done where two atoms from a dissociated molecule emitted an electron \cite{liu2015_double_slit_molecular_oxygen} or a photon \cite{alain_aspect_calcium2_recoil_interference_1982}, or where light (in form of a microwave field) was used as a beam splitter for atoms \cite{haroche2001complementarity}. We emphasized that although the fundamental concept of quantum complementarity can be demonstrated in all different configurations of scattering experiments (Fig. \ref{fig:interferometer}), there are major differences in the intermediate case where only partial information is revealed about the path of the photon. In the case of independent scatterers (slits), which-way information is recorded in the location of recoil excitation and not prone to dephasing, while for a single scatterer or rigidly coupled scatterers such information lies in the phase, and is not recorded when the scattering process is much slower than the confinement trap period.

It is amusing to note that the discussions between Einstein and Bohr focused on the the second configuration (Fig. 1C) which, as we have shown here, is a more complicated and subtle case compared to the independently movable slits (Fig. 1B), but they were mainly interested in the limiting cases of complementarity where  $|\beta|^2$ is either small or large.

\textit{Acknowledgments}.---We acknowledge support from the NSF (grant No. PHY-2208004), from the Center for Ultracold Atoms (an NSF Physics Frontiers Center, grant No. PHY-2317134), from the Vannevar-Bush Faculty Fellowship (grant no. N00014-23-1-2873), from the Gordon and Betty Moore Foundation GBMF ID \# 12405), from the Army Research Office (contract No. W911NF2410218) and from the Defense Advanced Research Projects Agency (award HR0011-23-2-0038). Yoo Kyung Lee acknowledge the MathWorks Science Fellowship. Yu-Kun Lu is supported by the NTT Research Fellowship.\\

\newpage
\clearpage

\setcounter{figure}{0}
\setcounter{equation}{0}
\makeatletter 
\renewcommand{\thefigure}{S\@arabic\c@figure}
\renewcommand{\theequation}{S\@arabic\c@equation}

\makeatother

\onecolumngrid


\begin{thebibliography}{18}
\bibitem [{foo()}]{footnote}%
  \BibitemOpen
  \href@noop {} {\bibinfo {title} {{We can in principle use the symmetric/antisymmetric states $\left(|0,1;1\rangle\pm|0,1;1\rangle\right)/\sqrt{2}$ as basis as well, and the detection of such states does not reveal which-way information. However, here we assume a small but non-zero potential energy offset between the two atoms to lift the degeneracy, so states $|0,1;1\rangle$ and $|1,0;1\rangle$ are the unique eigenstates of the system.}}}\BibitemShut {Stop}%
\end{thebibliography}
\end{document}